\documentclass[10pt,aps,prd,preprintnumbers,tightenlines,
   showpacs,nofootinbib]{revtex4-1}
\newcommand{\PRE}[1]{} 

\usepackage{hyperref}      
\usepackage{bm}
\usepackage{epsfig}
\usepackage{subfigure}

\def\beq{\begin{eqnarray}}
\def\eeq{\end{eqnarray}}
\def\bea{\begin{eqnarray}}
\def\eea{\end{eqnarray}}

\newcommand{\mweak}{m_W}

\newcommand{\OmegaDM}{\Omega_{\text{DM}}}

\newcommand{\gev}{\text{GeV}}
\newcommand{\tev}{\text{TeV}}

\newcommand{\eqref}[1]{Eq.~(\ref{#1})}

\newcommand{\Figref}[1]{Figure~\ref{fig:#1}}

\newcommand{\Omegachi}{\Omega_{\chi}}

\newcommand{\mhu}{m_{H_u}}

\begin{document}

\preprint{CALT-68.2941}

\PRE{\vspace*{0.7in}}

\title{ \PRE{\vspace*{0.3in}} Prospects of Focus Point Supersymmetry
  for Snowmass 2013}

\author{David Sanford
\PRE{\vspace*{.3in}}}
\email{dsanford@caltech.edu} 
\affiliation{California Institute of Technology, Pasadena, CA 91125,
  USA
\PRE{\vspace*{.5in}}} 

\begin{abstract}
\PRE{\vspace*{.2in}} We briefly review the motivations and features of
focus point supersymmetry and in particular the focus point region of
the CMSSM.  Applying the constraint that the neutralino is a thermal
relic, we examine current and projected collider and dark matter
constraints on the focus point region.  We demonstrate that the focus
point region is currently constrained by multiple dark matter
experiments, and future sensitivy on multiple fronts will probe large
portions of the parameter space.
\end{abstract}

\maketitle

One of the driving motivations for supersymmetry (SUSY) theories is
the hierarchy problem of the Standard Model (SM).  The presence of
superpartners at the weak scale cancels quadratic divergences in the
Higgs potential, drastically reducing the fine-tuning present in the
theory.  However, current LHC results generically imply squark masses
of $\gtrsim 1~\tev$.  In most SUSY models this implies reintroduction
of fine-tuning, which, while far less severe than the fine-tuning
present in the SM, weakens the motivation of SUSY theories.

This issue is compounded by the discovery of the Higgs boson at the
LHC with a mass of $m_h \approx
125.6~\gev$~\cite{Aad:2012tfa,Chatrchyan:2012ufa}.  In the minimal
supersymmetric standard model (MSSM), it is well-known that the
tree-level Higgs mass cannot exceed $m_Z \simeq 91~\gev$, but
radiative corrections raise the Higgs mass significantly.  However,
achieving a Higgs mass consistent with the experimental determination
requires either stop masses of $\mathcal{O}(8-10)~\tev$ or a large
stop $A$-term.  Large $A$-terms are non-generic, while stop masses of
$\mathcal{O}(8-10)~\tev$ reintroduce fine-tuning roughly two orders of
magnitude worse than implied by current collider limits.

Relatively heavy scalars are also motivated from precision observables
in the flavor and CP violation sectors.  General weak-scale SUSY
models suffer serious constraints from flavor violating observables,
and though well-known mechanisms exist to escape such effects their
severity is also reduced by increasing sfermions masses.  Moreover,
even in such mechanisms CP violation is generally present, and
motivates sfermion masses in the multi-TeV range to avoid electron and
neutron EDM constraints~\cite{Feng:2010ef}.

One framework which addresses this combination of issues is focus
point (FP) supersymmetry~\cite{Feng:1999mn,Feng:1999zg}.  In the MSSM
for $\tan\beta \gtrsim 5$, electroweak symmetry breaking requires
\begin{equation}
m_Z^2 \approx - 2 \mu^2 - 2 \mhu^2 (\mweak) \ ,
\label{mz2}
\end{equation}
at tree level.  If $m_Z^2 \sim \mu^2 \approx \left| \mhu^2 \right|$,
the theory is relatively natural, while it becomes fine-tuned when
$m_Z^2 \ll \mu^2 \approx \left| \mhu^2 \right|$.  The former condition
condition is of course satisfied if all SUSY-breaking masses are
$\mathcal{O}(M_{\mathrm{weak}})$, but it can also be satisfied if
SUSY-breaking masses are significantly larger than $m_Z$ at the
SUSY-breaking scale but $\mhu^2 \rightarrow 0$ at the weak scale.
This mechanism is present in FP SUSY models, where the Higgs potential
exhibits ``radiative naturalness'' due to renormalization group (RG)
runnning.

The required boundary conditions are dependent on the SUSY-breaking
scale, and for $M_{\mathrm{SUSY-breaking}} \sim M_{\mathrm{GUT}}$ one
solution is the case of unified scalar masses, zero $A$-terms, and
small gaugino masses.  This scenario is realized in the constrained
MSSM (CMSSM) with $A_0 = 0$ and $M_{1/2} \ll m_0$, producing the
``focus point region'' of the CMSSM where $m_0$ is in the multi-TeV
range but $\mu$ remains weak scale.  Fine-tuning in this region is
reduced by approximately an order of magnitude relative to generic
MSSM models with equivalent stop masses~\cite{Feng:2012jfa}.

While the collider prospects for the FP region are poor due to large
superpartner masses, large portions can be probed at a variety of dark
matter experiments.  Neutralino dark matter is perhaps the most
studied dark matter candidate, and it is well-known that the simplest
cases produce a thermal relic density either larger or smaller than
the observed dark matter abundance.  A pure Bino is generally
overabundant while a pure Wino or Higgsino is underabundant for
neutralino masses $m_\chi\lesssim1~\tev$, and some further mechanism
is required to bring the relic density in agreement with the observed
dark matter abundance.  One such mechanism is Bino-Higgsino mixing,
which is realized in the FP region due to the relatively small values
of $\mu$ required for reduced fine-tuning.  Significant Bino-Higgsino
mixing also enhances both direct detection and annihilation signals,
improving the ability of dark matter experiments to probe FP models.
Spin-independent direct detection is particularly interesting for such
models, with cross-sections within the reach of current and
near-future direct detection experiments~\cite{Feng:2010ef}.

In exploring the FP parameter space, a particularly interesting region
is the slice wherein the neutralino thermal relic density saturates
the observed relic density,
\begin{equation}
\Omegachi\left( m_0, M_{1/2}, A_0, \tan\beta,
\mathrm{sign}\left(\mu\right) \right) = \OmegaDM\ .
\end{equation}
By fixing the sign of $\mu$ and setting $A_0 = 0$ to generate the
desireable FP RG behavior, for a particular choice of $\left\{M_{1/2},
\tan\beta \right\}$ there exists a unique value of $m_0$ for which
$\Omegachi = \OmegaDM$~\cite{Feng:2011aa}.  This allows the FP region
to be studied in the $\left\{M_{1/2}, \tan\beta \right\}$ plane
wherein \textit{every point has the correct thermal relic density}.
Applying this cosmological constraint allows a larger range of
parameters to be studied.

\begin{figure}[tb]
  \subfigure[$m_0$ and $m_\chi$ for $\Omegachi \approx 0.23$.]{
    \includegraphics[width=0.48\columnwidth]{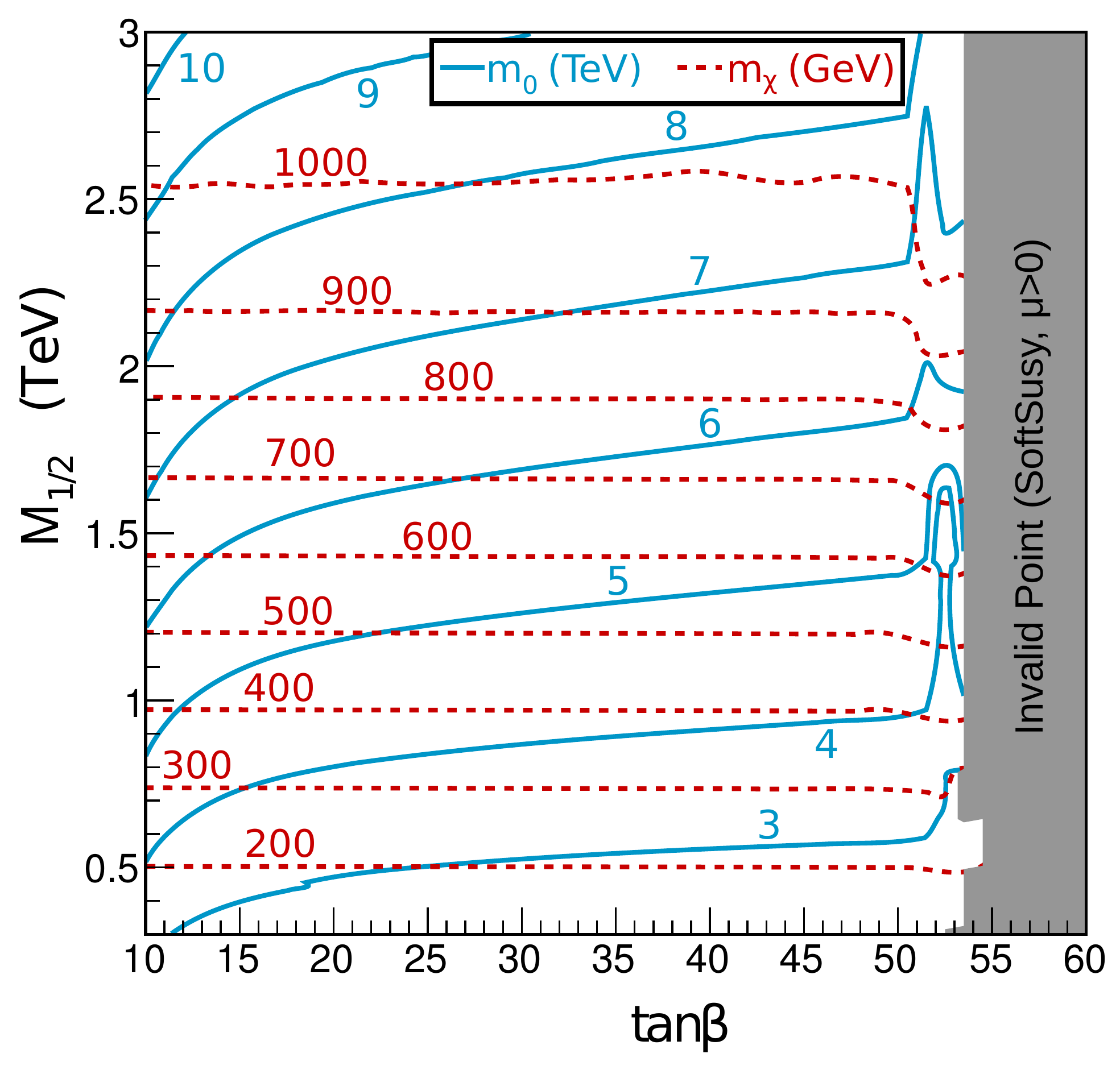}
    \label{fig:m0mchi} }
  \subfigure[Limits on FP SUSY for $\Omegachi \approx 0.23$.]{
    \includegraphics[width=.48\textwidth]{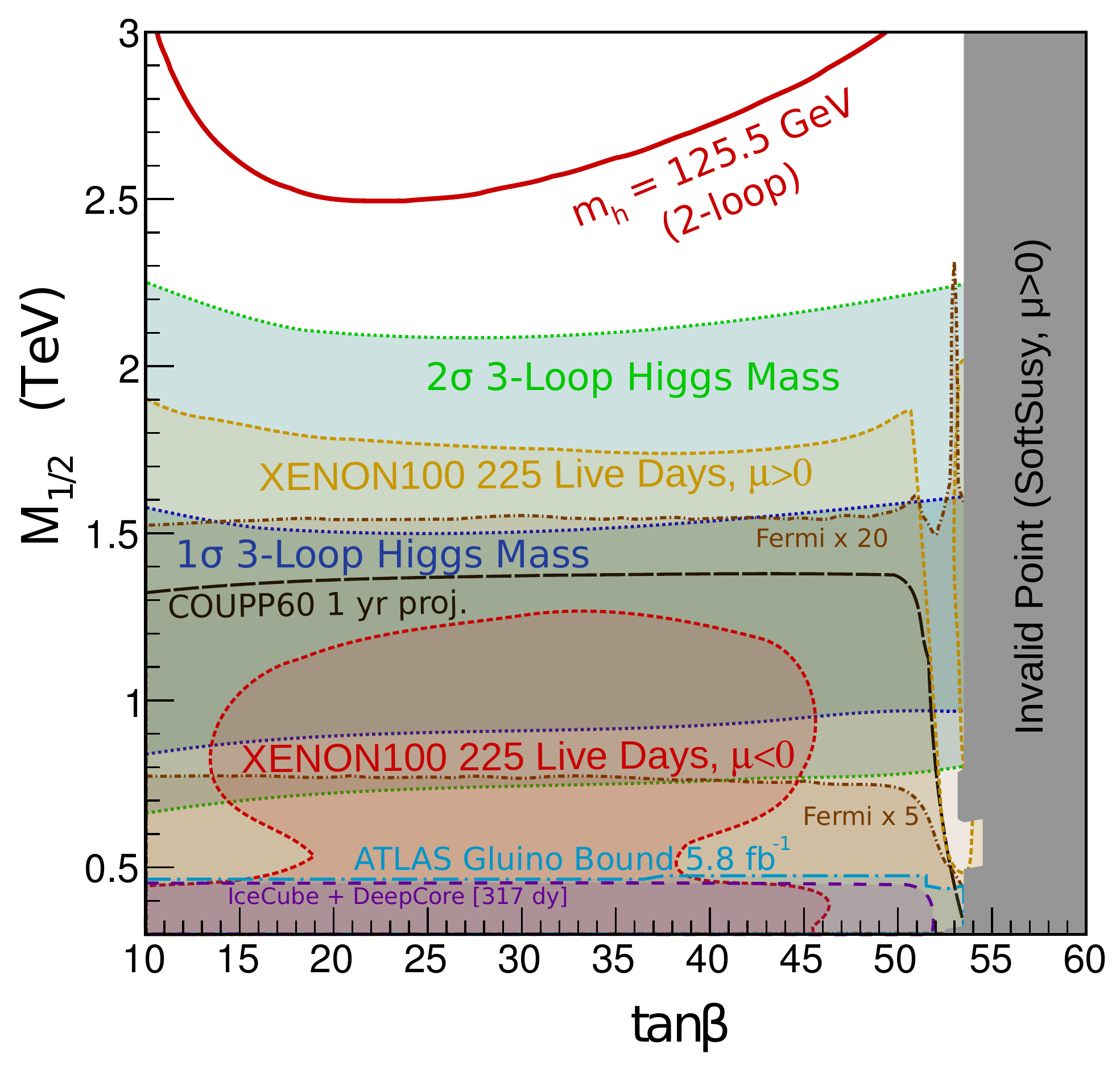}
    \label{fig:alllimits} }
\vspace*{-.1in}
\caption{\label{fig:fp} \textit{The Focus Point Region for $\Omegachi
    = \OmegaDM$.}  Shown are the values of $m_0$ and $m_\chi$
  satisfying the thermal relic density (left) and relevant constraints
  (right).  The collider constraint is extrapolated from ATLAS gluino
  searches using $5.8 fb^{-1}$ of data with
  $\sqrt{s}=8~\tev$~\cite{ATLAS:2012ona}.  Current dark matter
  constraints use results from XENON100 with 225 live
  days~\cite{Aprile:2012nq} and IceCube (w/ DeepCore) results using
  317 live days~\cite{:2012kia}.  Projected limits are drawn from
  projected COUPP60 sensitivity after a 1 year physics
  run~\cite{coupp60} and multiples of the current Fermi-LAT
  sensitivity from a stacked dwarf spheroidal analysis~\cite{lat}.
  The two-loop Higgs mass is determined by
  \textsc{SOFTSUSY}~\cite{Allanach:2001kg}, while the three-loop
  result uses \textsc{H3m}~\cite{Harlander:2008ju,Kant:2010tf}.  Plots
  are taken from Ref.~\cite{Draper:2013cka}.}
\end{figure}

Constraints on the FP region using this framework were studied in
Refs.~\cite{Feng:2011aa,Draper:2013cka}, the latter of which (and
plots here) used \textsc{SOFTSUSY}~3.1.7~\cite{Allanach:2001kg} for
spectrum generation, \textsc{MicrOMEGAs}~2.4~\cite{Belanger:2010gh} to
calculate the relic density and direct detection processes, and
\textsc{DarkSUSY}~5.0.5~\cite{ds} to calculate indirect detection
rates.  Results are shown in \Figref{fp}, using the relic density
determination of WMAP with 7 years of data~\cite{Komatsu:2010fb}.
\Figref{m0mchi} shows the value of $m_0$ required to achieve
$\Omegachi = \OmegaDM$ and the corresponding neutralino mass $m_\chi$.
Generally the required value of $m_0$ increases with increasing
$M_{1/2}$ and decreases with increasing $\tan\beta$.  However, for
$\tan\beta \gtrsim 50$ the appropriate value shifts downward, as the
pseudoscalar Higgs bosons becomes light enough in this region to
signficantly alter the relic density calculation.

\Figref{alllimits} shows associated constraints upon the FP region.
Due to the relatively high squark mass scale, the dominant constraints
are derived from searches for gluino
pair-production~\cite{ATLAS:2012ona}.  The associated reach is
relatively limited, constraining $M_{1/2} \lesssim 500~\gev$ almost
independent of $\tan\beta$.  Even with current data, direct detection
experiments place significantly stronger bounds, with current
XENON100~\cite{Aprile:2012nq} results constraining $M_{1/2} \gtrsim
1.8~\tev$ for nearly the entire range of $\tan\beta$ shown if $\mu >
0$, with somewhat stronger bounds at large and small $\tan\beta$.  For
$\mu < 0$ the constraints are weaker, requiring $M_{1/2} \gtrsim
1~\tev$ for a moderate range of $\tan\beta$ but weakening to $M_{1/2}
\gtrsim 500~\gev$ for small $\tan\beta$ and placing no constraint for
$\tan\beta \gtrsim 45$.  Spin-independent limits were produced using a
strange quark form factor of $f_s =
0.05$~\cite{Freeman:2009pu,Young:2009zb,Giedt:2009mr}.  While current
spin-dependent results do not have sensitivity to the focus point
parameter space, near future results from COUPP60~\cite{coupp60} are
expected to constrain $M_{1/2} \lesssim 1.3-1.4~\tev$ for $\tan\beta
\lesssim 50$.  Indirect detection experiments are also relevant, with
current IceCube~\cite{:2012kia} results constraining $M_{1/2} \gtrsim
500~\gev$ for $\tan\beta \lesssim 50$, producing a bound competitive
with current LHC constraints.  Moreover, while gamma ray searches
currently do not probe the FP region, an order of magnitude
improvement on current Fermi-LAT sensitivity from dwarf
spheroidals~\cite{lat} will provide significant sensitivity to
$M_{1/2} \sim 700~\gev - 1.5~\tev$.  The generation of IceCube and
Fermi-LAT bounds are detailed in Ref.~\cite{Draper:2013cka}.  Future
experiments on all these fronts will have the ability to probe
significantly larger regions of FP parameter space.

These sensitivities are especially important given recent computations
of the Higgs mass with leading 3-loop effects included in models with
multi-TeV scalars~\cite{Draper:2013cka,Feng:2013tvd}.  Generally
2-loop determinations require stop masses of $\mathcal{O}(8-10)~\tev$,
which in the FP region with $A_0 = 0$ places the appropriate Higgs
mass at $M_{1/2} \gtrsim 2.5~\tev$, beyond the reach of current or
near-future experiments.  However, including 3-loop effects reduces
the required stop mass to $3-4~\tev$ even without left-right mixing,
improving the prospect of dark matter and collider experiments to
probe FP models, with a favored region of $700~\gev \lesssim M_{1/2}
\lesssim 2.2~\tev$~\cite{Draper:2013cka}.

\section*{Acknowledgments}

We are grateful to P.~Draper, J.~Feng, P.~Kant, and S.~Profumo for
collaboration relevant to this work, and K.~Matchev for useful input
in preparing this manuscript.  DS is supported in part by
U.S.~Department of Energy grant DE--FG02--92ER40701 and by the Gordon
and Betty Moore Foundation through Grant No.~776 to the Caltech Moore
Center for Theoretical Cosmology and Physics.

\bibliography{bibfpsnowmass}{}

\end{document}